\documentclass{jfm}
\usepackage{graphicx}
\graphicspath{{figures/}}
\usepackage{epstopdf, epsfig}

\usepackage{todonotes}
\usepackage{dcolumn}
\usepackage{bm}
\usepackage{amsmath}
\usepackage{gensymb}

\usepackage{pifont}
\newcommand{\perB}{%
  \resizebox{!}{1.2\fontcharht\font`0}{$\mkern-2mu\times\mkern-2mu$}%
}

\shorttitle{Capillary Ripples}
\shortauthor{M. Jalaal, C. Seyfert and J. H. Snoeijer}

\title{Ripples in Thin Films}

\author{Maziyar Jalaal
  \corresp{\email{mazi@alumni.ubc.ca, c.seyfert@utwente.nl}},
  Carola Seyfert
 \and Jacco H. Snoeijer}

\affiliation{\aff{1}Physics of Fluids Group, Max-Planck-Center for Complex Fluid Dynamics, \\ University of Twente, Drienerlolaan 5, 7522NB Enschede, The Netherlands}

\begin{document}

\maketitle

\begin{abstract}
Capillary ripples on thin viscous films are important features of coating and lubrication flows. Here we present experiments based on Digital Holographic Microscopy, measuring the morphology of capillary ripples ahead of a viscous drop spreading on a prewetted surface \textcolor{black}{with a nanoscale resolution.} Our experiments reveal that upon increasing the spreading velocity, the amplitude of the ripples first increases and subsequently decreases. Above a critical  spreading velocity, the ripples even disappear completely and this transition is accompanied by a divergence of the ripple wavelength.  These observations are explained quantitatively using linear wave analysis, beyond the usual lubrication approximation, illustrating that new phenomena arise when the capillary number becomes order unity.  
\end{abstract}


\section{Introduction}


Visco-capillary ripples are common features of free surface flows involving thin viscous films. A prime example of such ripples is given by long bubbles moving in a narrow tube that are separated from the wall by a thin lubrication layer, known as Bretherton bubbles \citep{Bretherton1961}. While the interface shape is monotonous at the bubble\textsc{\char13}s advancing side, the thin film at the rear exhibits steady capillary waves, that are co-moving with the bubble. 
The ripples are found in many other flow phenomena, for example in dip-coating processes, for both plunging and withdrawing plates \textcolor{black}{\citep{Wilson1982, Snoeijer2008, Maleki2011}}, spreading droplets on a pre-wetted film \citep{tanner1979spreading,Tuck1990,Cormier2012,Bergemann2018,teisala2018wetting}, dewetting fronts \citep{fetzer2006slip, snoeijer2010asymptotic}, and levelling of capillary films \citep{mcgraw2012self, Salez2012, Ilton2016}. Apart from their intrinsic interest, the ripples are important for establishing boundary conditions in coating problems \citep{Bretherton1961,Taroni2012, Giavedoni1999}, and give rise to intricate bifurcation scenarios for the forced wetting transition \citep{Ziegler2009,kopf2014emergence,Lin2016}. In the context of dewetting polymer films, the shape of a ``dimple" was even used to quantify the presence of slip \citep{Jacobs1998, Fetzer2007, Baumchen2010}, and thus serve as a tool to infer material properties.



The emergence of capillary ripples can be understood from the lubrication equation for thin films. The gradient of capillary pressure, driving the flow, involves the derivative of the curvature, and hence third order spatial derivatives of the free surface profile. These high derivatives are responsible for the ripples, as they lead to modes of complex wave-numbers and hence (exponentially damped) oscillatory shapes. Indeed, ripples do not appear in gravity-driven flows, for which the pressure gradient involves only the first spatial derivative of the film profile \citep{huppert1982propagation}. 
As an illustrative example, we quote here the levelling of a sharp step in film thickness. The gravitational levelling is described by a diffusion-like equation that broadens the step along monotonic profiles, whereas capillary levelling gives rise to undulating profiles \citep{mcgraw2012self}. 

\begin{figure}
  \centerline{\includegraphics[scale=0.33]{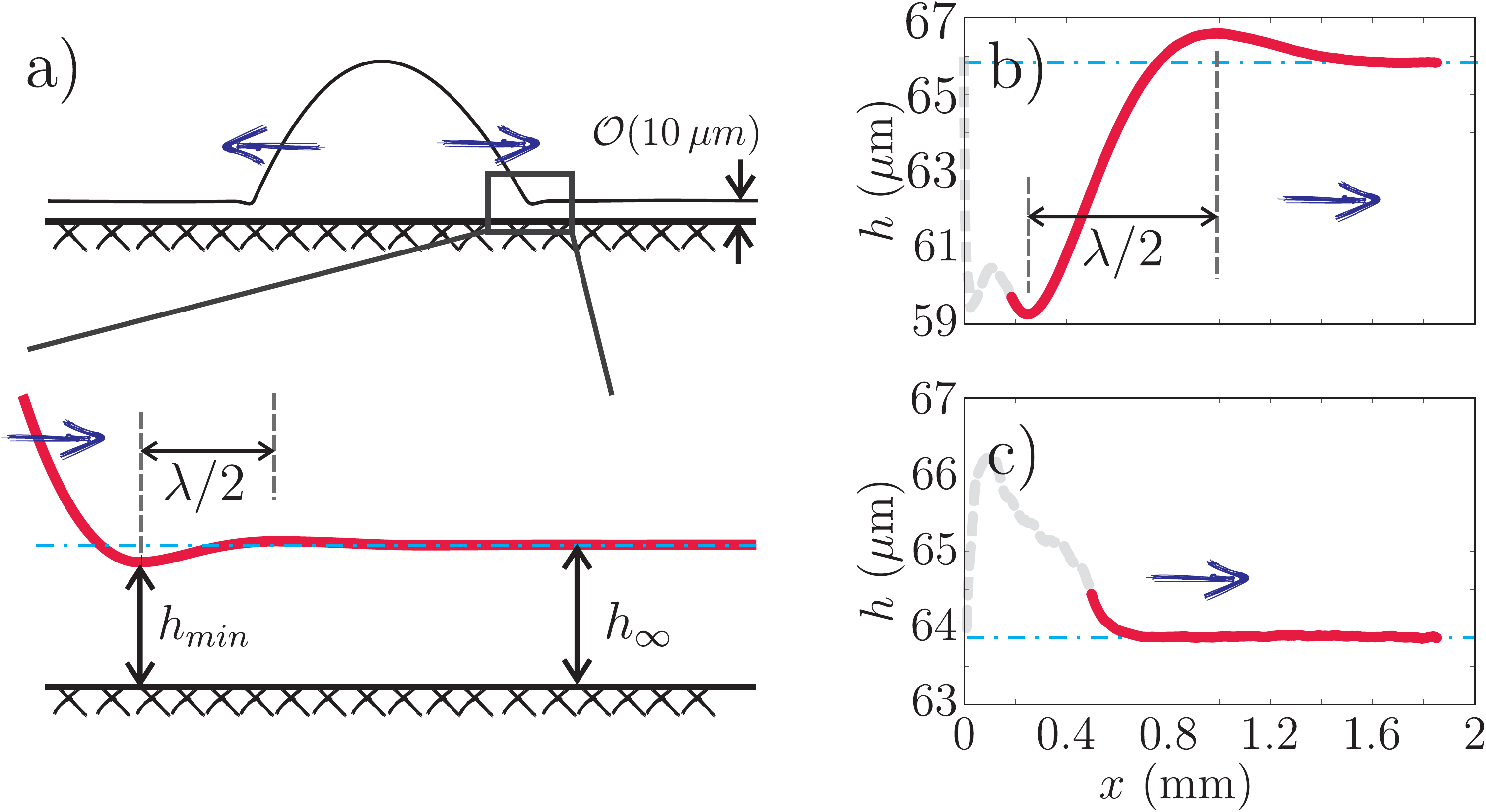}}
  \caption{
\textcolor{black}{  
  a) Sketch of the problem: a viscous drop is placed on a thin film of the same fluid and spreads (dimensions not to scale). The magnified view corresponds to a numerical solution based on the lubrication equation. 
b,c) Examples of experimentally obtained profiles, for $Ca = 0.0125$, where a distinct capillary ripple is observed (b), and  for $Ca = 0.3562$, where no ripple occurs (c). The red line indicates the interface profile that can be resolved by Digital Holographic Microscopy \ref{sec:MM}). In all panels, the arrow shows the direction of the motion and the blue dashed-dotted line indicates the pre-wetted film thickness.}
  }
\label{fig:intro}
\end{figure}

The theoretical coverage of capillary ripples is extensive, but several of the presented scenarios have yet to be confirmed experimentally. In this paper we experimentally study the prototypical example of a viscous drop spreading over a film of the same liquid (see figure \ref{fig:intro}a). 
Using Digital Holographic Microscopy (DHM), we quantify the visco-capillary ripples close to the \textit{``contact line''}, where the front of the droplet connects to the pre-wetted film. Typical experimental results are shown in panels b and c of figure \ref{fig:intro}.
The analysis of this problem using lubrication theory goes back to pioneering works of \citet{tanner1979spreading} and \citet{Tuck1990}. In the limit where the drops are asymptotically large compared to the film thickness, the dimple is predicted to exhibit a universal profile (inset Figure 1a) which also plays a key role in the description of dewetting fronts \citep{snoeijer2010asymptotic}. An interesting characteristic of this solution is that the minimal film thickness $h_{min}$, compared to the film thickness $h_\infty$, exhibits a significant depression $h_{min} / h_{\infty} \approx 0.82$, as reported by \cite{tanner1979spreading,Tuck1990}.

Here we show that the dimples can be less pronounced than the asymptotic value of $h_{min} / h_{\infty} \approx 0.82$ (e.g. see Figure \ref{fig:intro}b), and exhibit a nontrivial dependence on spreading velocity. Remarkably, the oscillations even disappear both at low and at high spreading velocities: the ripples give way to a monotonic approach of the prewetted film (Figure \ref{fig:intro}c). The experiments are interpreted using a linear wave analysis, closely following theoretical work on generic coating flows by \citet{Taroni2012}. At high velocity, it is shown that ripples disappear due to the effect of large capillary numbers, i.e beyond the lubrication hypothesis. The wavelengths are found to diverge at a critical speed, in agreement with theory. At low velocity, we discuss the effect of gravity and the importance of separation of length scales.

\section{Experiments}

\subsection{Materials \& Methods}


\label{sec:MM}

\begin{figure}
  \centerline{\includegraphics[width=13cm]{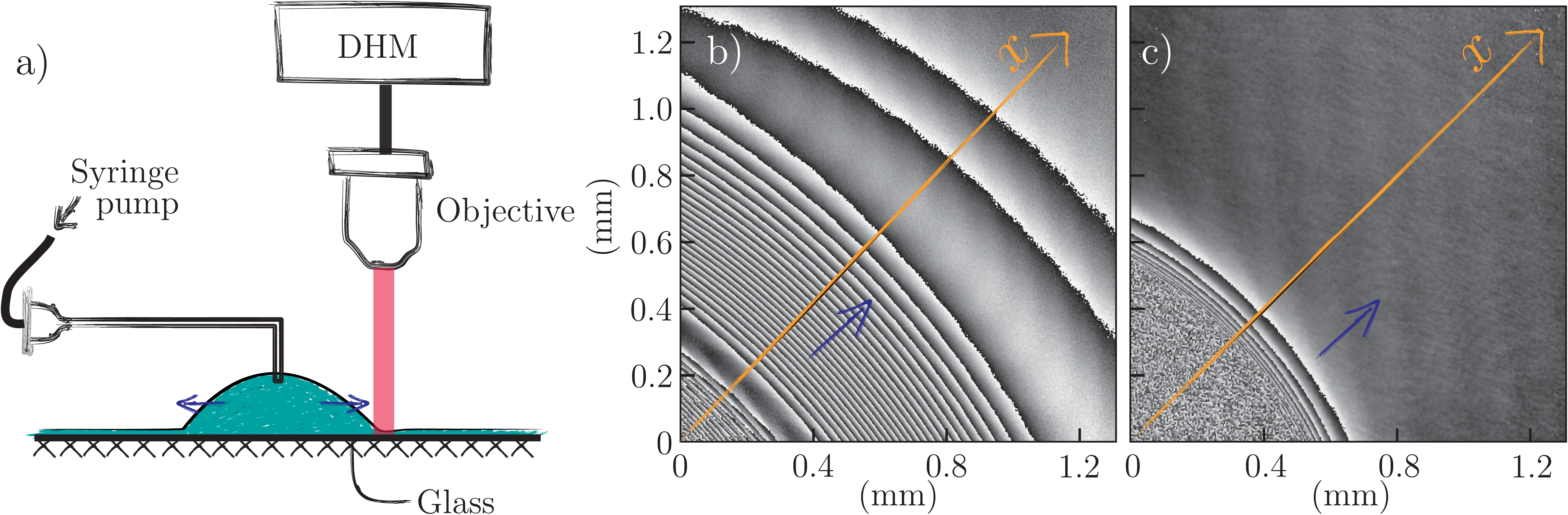}}
  \caption{
\textcolor{black}{  
  a) The experimental setup for the investigations into the undulating surface of a thin film at the edge of a moving droplet. 
  Droplets were gently placed (or extruded) on a thin layer of the same liquid. Holograms are captured from above using a reflection DHM.
  Examples of phase images are shown for b) $Ca = 0.0125$ and c) $Ca = 0.3562$. 
  In panels b and c, the diagonal arrows indicate the direction of the motion. 
  Along this diagonal, the height profiles seen in figure \ref{fig:intro} are extracted. 
  The phase images show noisy parts in the lower left corners, where the surface slope of the droplet is too steep to be resolved. These parts correspond to the dashed lines in the profiles of figure \ref{fig:intro}. The blue arrows indicate the direction of spreading.
  }
  }
\label{fig:setup}
\end{figure}

A schematic of the experimental setup is given in Figure \ref{fig:setup}a. Using a needle, we deposit an oil droplet on a pre-wetted oil surface.
The fluid used in the experiments is a silicone oil of viscosity \textcolor{black}{$\eta=$12.125 Pa$\cdot$s. }The surface tension of the material is $\gamma = 0.021\,$N/m
and the density is $\rho=972\,$kg/m$^3$. Thin films of different thicknesses in the range of $15\, \mu$m $\leq h_{\infty} \leq 65\, \mu$m were acquired by spin-coating the silicone oil on cleaned glass slides. The height of the films was measured with a spectroscope at multiple locations to ensure the flatness.
We could control the spreading velocity by changing the rate of extrusion of the syringe pump (Harvard PHD ULTRA\textsuperscript{TM} 70-3006) connected to the needle. 
When extruding, we ensure that the droplets are sufficiently large so that the extrusion process had no influence on the spreading dynamics. In those cases, the height of the droplets is in the same order of magnitude as the capillary length, $\ell_\gamma = \sqrt{\gamma/\rho g} \sim \mathcal{O}$(1\,mm), and much larger than the film thickness.
To achieve low spreading velocities, we deposited a droplet manually or with the syringe pump on the film. Without further extrusion, the droplets spread spontaneously because of the capillary action as well as their weight. Due to the high viscosity of the chosen silicone oil, some of the experiments could take more than 30 hours. 
In the latter protocol, at very small velocities, the height of the droplet can become comparable to the height of the pre-wetted film.
The spreading velocity, $v$, is measured by dividing the distance the first extremum travelled between frames by the time passed between the frames. Once the ripples disappeared, we opted to measure the front velocity by setting the travelling point at the first measured height line of the phase image (e.g. see panel c of figure \ref{fig:setup}). 

To measure the ripples, we imaged the droplets from above, where the front of the droplet invades a fixed field of view. Our main tool of measurement was a reflection digital holographic microscope (LynceeTec DHM\textsuperscript{\textregistered}-R1000). \textcolor{black}{The device records holograms, created through laser interference, with a CCD camera (Mikrotron EoSens 4CXP). From these holograms, one can numerically reconstruct intensity and phase information.
Herein, the phase information provides the height profile of the thin film. Panels b and c in figure \ref{fig:setup} show phase images corresponding to the reconstructed height profiles shown in panels b and c of figure \ref{fig:intro}.} 
A great advantage of DHM is the non-invasive, non-scanning observation of a sample and the high resolution in line with the laser beam. 
%
%
For a more detailed description of DHM, see e.g. \citet{Kim2010}. In our system, the vertical resolution in line with the laser beam is $\sim$1\,nm. The lateral resolution is limited by the magnification of the used objective lens, but ranges between 2.85 $\mu$m (5\perB \, magnification) and 8.5 $\mu$m (1.25\perB \, magnification). We record the holograms from 5 frames per second down to 1 frame every 15 minutes. Note, that the slope of the surface, which the reflection DHM can resolve, depends on the numerical aperture of the used lens. For the data shown here, the maximum slope varies in between 3.4\degree (5\perB \,  magnification) and 1.1\degree (1.25\perB \, magnification). Steeper slopes will result in a noisy phase signal. \textcolor{black}{This can be seen from the speckle-like patterns in panels b and c in figure \ref{fig:setup}, which are the data corresponding to the dashed-grey part of profiles in panels b and c of  figure \ref{fig:intro}.} 
\subsection{Amplitude and wavelength of the ripples}


We extract two main features from the experimentally measured height profiles: the minimum height, $h_{min}$ and the ripple wavelength, $\lambda$. Note that the waves are exponentially damped, so that the position of the second minimum is difficult to determine experimentally. We therefore measure half of the wavelength, $\lambda/2$, from the distance between the minimum and the first maximum (see figure \ref{fig:intro}b). Our prime interest is to investigate how $h_{min}$ and $\lambda$ evolve as functions of the capillary number $Ca =\eta \,  v / \gamma$.

\begin{figure}
  \centerline{\includegraphics[width=13cm]{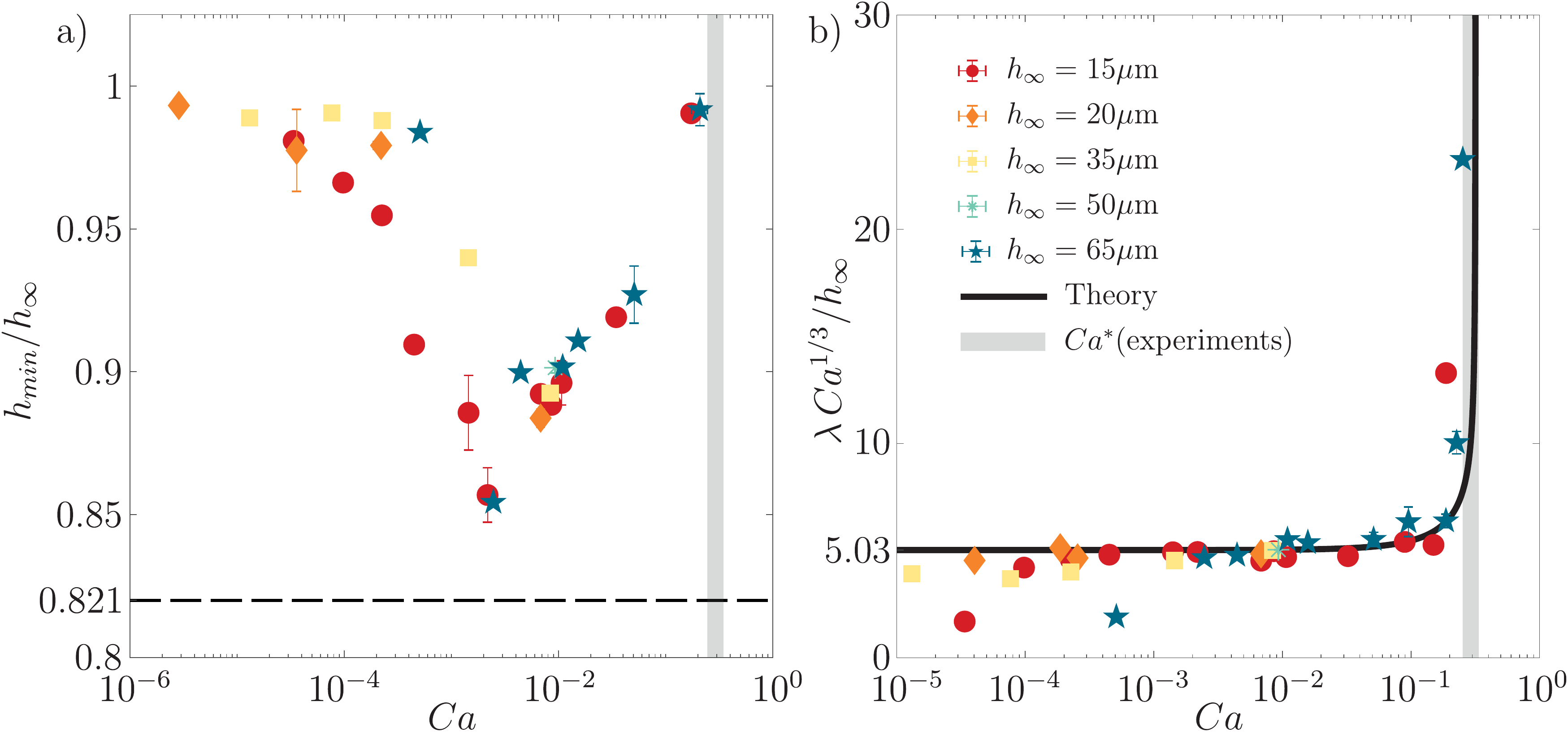}}
  \caption{a) The normalized value of the minimal film thickness, $h_{min}/h_{\infty}$ and b) the normalized wavelength, as a function of the capillary number $Ca=\eta \, v/\gamma$. Different symbols correspond to different film thickness $h_\infty$ (see legend). The vertical grey regions show the critical $Ca^*$ beyond which dimples are not observed. The dashed grey line in panel (a) indicates the asymptotic prediction of lubrication theory without gravity. The solid line in panel (b) is the prediction of equation \ref{equ:Disp}. }
\label{fig:ExpRes}
\end{figure}

Figure \ref{fig:ExpRes}a shows the normalised film thickness, $h_{min}/h_\infty$ versus the capillary number. Different symbols correspond to different film thickness, $h_\infty$. 
The minimum film height in our experiments comes close to the predicted asymptotic value of $0.82 \, h_\infty$, but stays above that for all $Ca$.
The deepest dimples roughly correspond to $h_{min}/h_\infty \approx 0.85$ around $Ca \approx 2 \times 10^{-3}$. Intriguingly however, the dimple height exhibits a non-monotonic dependence with $Ca$. The ripples tend to flatten for both increasing and decreasing spreading velocities, with $h_{min}/h_\infty \rightarrow 1$. This trend is observed for all film thicknesses. 

A remarkable feature appears at large speeds. Namely, we can clearly identify a critical capillary number, \textcolor{black}{$Ca \approx 0.295 \pm 0.042$} (the gray zone in figure \ref{fig:ExpRes}) at which the undulations suddenly disappear. At larger capillary numbers the droplet monotonically joins the viscous film, and no ripples are observed. This transition can be also be observed from the profiles in Figure \ref{fig:intro}b, c.

The evolution of the wavelength is reported in Figure \ref{fig:ExpRes}b. The prediction based on linear analysis of lubrication theory without gravity can for example be found in \cite{Bretherton1961}, and gives

\begin{equation}\label{eq:lambdalubri}
\lambda = \frac{4 \,\pi \, h_\infty}{3^{5/6} \, {Ca^{1/3}}}  \approx 5.03 \frac{h_\infty}{{Ca}^{1/3}}.
\end{equation}
We therefore report our measurement of the wavelength according to $\lambda \, Ca^{1/3} / h_{\infty}$, as a function of $Ca$. For small $Ca$, the rescaled wavelength indeed reaches a constant value $\approx 5$, in excellent agreement with (\ref{eq:lambdalubri}). Interestingly, the wavelength $\lambda$ suddenly  diverges when $Ca$ approaches the critical value reported above, at the point where the dimples disappear. Clearly, this feature is not covered by the lubrication prediction.


%

\section{Linear analysis of the ripples}

\subsection{Interpretation}


We now interpret these experimental observations from the equations of Stokes flow. 
First, we remark that $h_{min}$ cannot be  predicted in a closed-form analytical expression. For example, the value $h_{min}/h_{\infty} \approx 0.82$ was obtained by numerical solution to the lubrication equation, matching to a large drop in the absence of gravity. However, such a lubrication calculation is clearly not applicable at large speeds, where $Ca$ approaches values of order unity. Hence, predicting $h_{min}$ will require full numerical resolution of Stokes flow. 

By contrast, explicit predictions for the wavelength $\lambda$ can be obtained from a linear analysis of the ripples, for arbitrary $Ca$. Such a linear analysis is indeed suitable, given that the ripples decay exponentially and thus have a small amplitude. When applied to lubrication theory without gravity this leads to (\ref{eq:lambdalubri}). The linear analysis for finite $Ca$, without gravity, was carried out by \citet{Taroni2012} for generic free surface inlet and outlet problems in coating flows. In fact, \citet{Taroni2012} identified a critical capillary number ${Ca}^*=1/\pi\approx 0.318$ in the dispersion relation -- which coincides with the critical values of $Ca$, found in our experiments (the grey regions in figure \ref{fig:ExpRes}). As we will show below, their analysis is indeed applicable to our experiments. We furthermore introduce gravity, to study the possible effects on the ripples at low $Ca$. Indeed, owing to the scaling $\lambda \sim {Ca}^{-1/3}$ implied by (\ref{eq:lambdalubri}), the wavelengths can approach the capillary length, $\ell_\gamma$ and gravity will start to play a role. The crossover to gravitational effects is expected when $\lambda \sim \ell_\gamma$, which using (\ref{eq:lambdalubri}) can be inferred from

\begin{equation}\label{eq:bondestimate}
\frac{\lambda^2}{\ell_\gamma^2} \sim \frac{Bo}{Ca^{2/3}},
\end{equation}
%
%
where, $Bo=\rho \,  g \,  h_{\infty}^{2}/\gamma$ is the Bond number. Indeed, the ratio $\lambda / \ell_\gamma$ can approach unity for the lowest spreading velocities in our experiment. Therefore, we below discuss the wavelengths of the ripples, for arbitrary $Ca$ and $Bo$. 

\subsection{The ripple wavelength at finite ${Ca}$ \& $Bo$}

We now compute the ripple wavelength from the perturbation analysis on the Stokes equation, including gravity. Introducing the streamfunction $\psi$, and the generalised pressure $\bar p = p+\rho \, g \, y$, we can split the problem, using the standard approach, to
\begin{equation}
	 \nabla^{4}\, \psi = 0, \quad \mathrm{and} \quad \nabla^2 \, \bar p = 0.
  \label{equ:Stokes}
\end{equation}
We perturb around the flat state $h=h_\infty$, which in the frame co-moving with the wave, implies $\psi=-v \,y$, and $\bar p=0$. The expansion then takes the form,
\begin{eqnarray}
h(x) &=& h_\infty + \epsilon \, e^{-\sigma x}, \quad  
\psi(x,y) = - v \,  y + \epsilon \, f(y)  \,e^{-\sigma x}, \quad
\bar p(x,y) = \epsilon \, q(y) \, e^{-\sigma  x},
\end{eqnarray}
%
%
and subsequently (\ref{equ:Stokes}) implies
\begin{equation}
f = A \, \cos (\sigma y) +B \, \sin (\sigma y) + C\, y \, \cos (\sigma y) + D \, y \, \sin (\sigma y), \quad q = E \, \cos (\sigma y) + F \, \sin (\sigma y).
\label{eq:fandq}
\end{equation}
Note that imposing Stokes equation $\nabla \, \bar p = \mu \, \nabla^2 \mathbf u$, one can express $\{E,F\}$ in terms of $\{C,D \}$, so that (\ref{eq:fandq}) only contains four independent constants. 

At the solid boundary we impose a no-slip boundary condition $\mathbf u = -v \, \mathbf e_x$, while at the free surface $\mathbf u \cdot \mathbf n=0$. The dynamic boundary condition at the free surface reads, $\boldsymbol{\tau}\cdot \mathbf n = -\gamma \, \kappa \, \mathbf n$, where $\kappa$ is the interface curvature. Finally, the stress can be written as $\boldsymbol{\tau} =  - (\bar p - \rho \, g \, y)+ \mu (\nabla \mathbf u + (\nabla \mathbf u)^T)$. These are five boundary conditions, which, after linearisation, constitute a problem for the four independent coefficients of (\ref{eq:fandq}). 
A nontrivial solution arises when satisfying the conditions: 

\begin{equation} \label{equ:Disp}
2 \, Ca \left(\cos^2\bar\sigma -  \bar \sigma^2 \right) +(1- \bar \sigma^{-2}Bo)\left(\bar \sigma - \sin \bar \sigma \cos \bar \sigma \right) =0,
\end{equation}
where we introduced $\bar \sigma=\sigma \, h_\infty$.
%
For $Bo=0$, one recovers the result by \citet{Taroni2012}. 


The wavelength of the ripples is encoded in the roots $\bar \sigma$ of equation \ref{equ:Disp}. The imaginary part ${\rm Im}(\bar \sigma)$ provide the wavenumber of the ripple, while the real part ${\rm Re}(\bar \sigma)>0$ describes the decay towards the pre-wetted film. We therefore visualise the location of the roots in the complex plane, and track their ``migration" upon varying the parameters $Ca$ and $Bo$. 

\begin{figure}
  \centerline{\includegraphics[width=10cm]{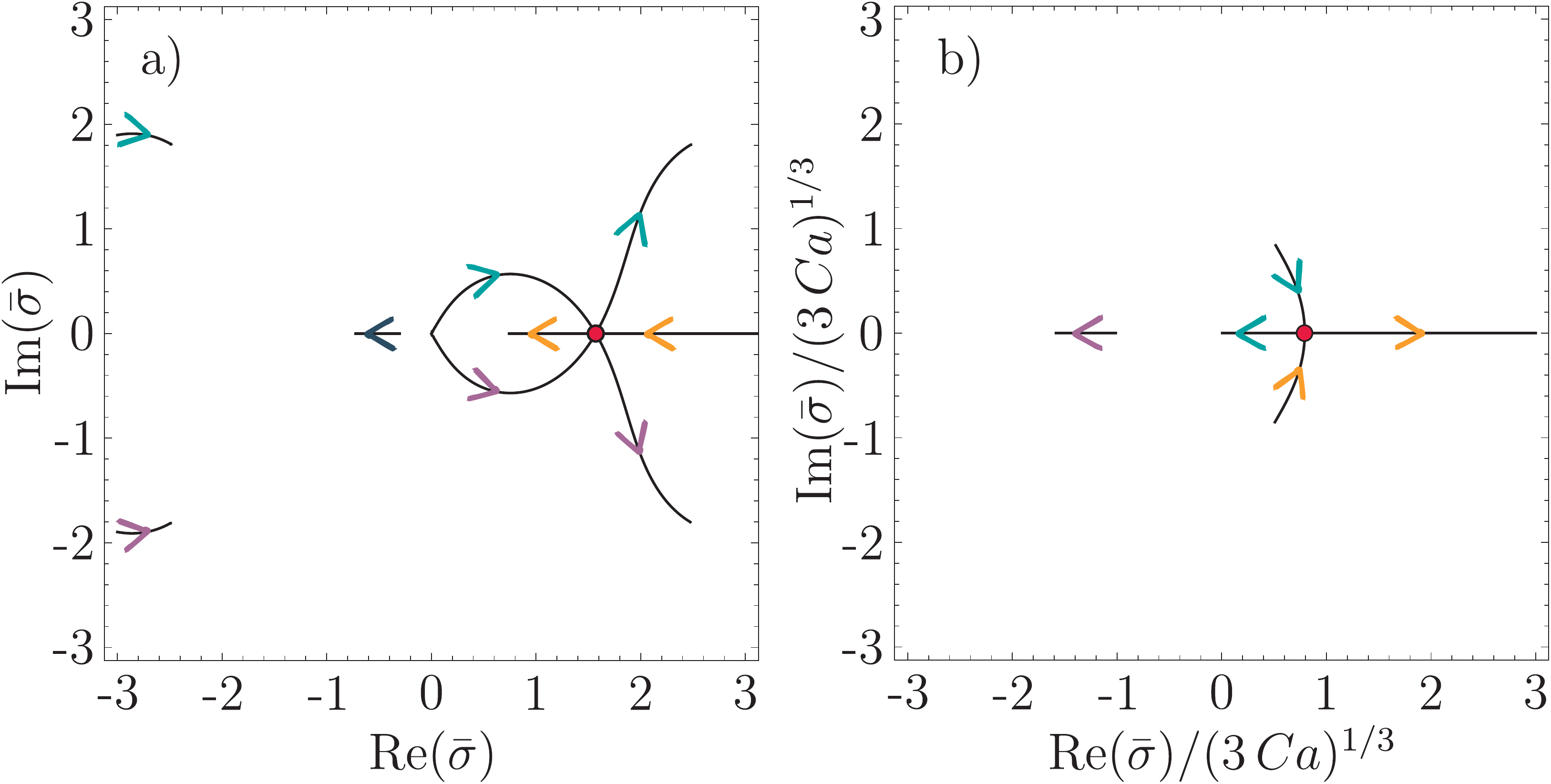}}
  \caption{a) The path of complex roots of equation \ref{equ:Disp}, where $Bo=0$. Arrows show the migration of the roots by increasing the values of $Ca$. The red circle corresponds to $Ca=Ca^*$. b) The path of complex roots in the lubrication limit, from equation \ref{equ:LubLimit}. 
  Arrows show the migration of the roots by increasing the values of $Bo$. The red circles denote $Bo=Bo^*$. 
 }
\label{fig:roots}
\end{figure}

Figure \ref{fig:roots}a shows the roots for increasing $Ca$ when $Bo=0$. At small $Ca$, the roots behave as expected from the lubrication limit. Three roots emanate from the origin, two of which have ${\rm Re}(\bar \sigma)>0$. These provide the oscillations as given by equation \ref{eq:lambdalubri}. At larger $Ca$, however, the two roots with ${\rm Re}(\bar \sigma)>0$ return to the real axis and merge at the point indicated by the red circle in figure \ref{fig:roots}a. 
This occurs at a critical capillary number $Ca^*=1/\pi$, at which ${\rm Im}(\bar \sigma)=0$. For $Ca > Ca^*$ the purely real root dominates, which means that a solution without the undulation remains. 
As the critical $Ca^*$ is approached from below, the wavelength diverges asymptotically as  $\lambda/h_{\infty} \sim 1/(Ca^* - Ca)^{1/3}$, as already reported by \citet{Taroni2012}. 
In fact, the prediction (\ref{equ:Disp}) provides an excellent description of the experimental data for the ripple wavelength (figure~\ref{fig:ExpRes}b, solid line). The diverging wavelength indeed coincides with $h_{min}/h_0 \rightarrow 1$ observed experimentally. Hence, the disappearance of the ripples at $Ca^*$ is due to appearance a critical point -- a feature that is not present in lubrication theory.


We now turn to the case of small $Ca$, for which gravity starts to play a role. In this regime, $|\bar \sigma| \ll 1$, and equation \ref{equ:Disp} reduces to 
\begin{equation}
\bar \sigma^3 = -3 \, Ca + \bar \sigma \, Bo,
\label{equ:LubLimit}
\end{equation}
which is essentially the lubrication limit. There are now three roots for all parameters. Figure \ref{fig:roots}b shows the migration of these roots as we increase the Bond number (the axes are rescaled according to 
equation \ref{eq:bondestimate}). Similar to what was previously observed, the two relevant roots migrate towards the real axis, as $Bo$ is increased. The two ripple-solutions disappear via a pitchfork bifurcation at $Bo^* = 2^{-(2/3)} \, Ca^{2/3}  \approx 0.63 \, Ca^{2/3}$, indicated by the red circle in figure \ref{fig:roots}b. This predicts again a diverging wavelength, now in the very small $Ca$ regime, for which $Bo^*$ becomes sufficiently small. The scaling near the critical point is $\lambda/h_{\infty} \sim Ca^{1/3}/(Bo^* -Bo )^{1/2}$, as is typical near a pitchfork bifurcation. 

While the experiments at small $Ca$ reveal $h_{min}/h_0 \rightarrow 1$, and hence a disappearance of the ripples, the diverging wavelength is not observed. Indeed, for the experimental values, this gravitational transition would require $Ca\sim \mathcal{O}(10^{-7})$, which is 1 or 2  orders of magnitude smaller than we were able to reach. Hence, in our experiments the waves cannot disappear due to gravity -- instead, we suspect that $h_{min}/h_0 \rightarrow 1$ is a finite size effect. Namely, the low velocities are reached in the final stages of the spreading experiments, where the height of the droplet becomes comparable with the thickness of the film. Hence we loose the separation of scales between the inner (``contact line'') and outer (droplet) regions. This lack of scale-separation also provides an explanation why our experiments do approach the asymptotic value $h_{\min}/h_\infty\approx 0.82$ predicted by \cite{Tuck1990}, but remain always slightly above. 

\section{Conclusions}

In this paper, we systematically explored the shape of capillary ripples in a thin oil film when a droplet of the same liquid spreads over the surface. Using a reflective digital holographic microscope, we could measure the micro-scale features of the thin film. Unexpectedly, we found that amplitude of the ripple increases, reaches a maximum around $Ca\approx 2 \times 10^{-3}$, and then decreases at larger capillary numbers. Above $Ca \approx0.3$ the dimples completely disappear, in excellent agreement with the linear analysis by \citet{Taroni2012}, predicting a critical capillary number $Ca^* = 1/\pi$. Atypically for thin film flows, these features cannot be explained by the lubrication approximation. Since surface tension is responsible for the ripples, one can on a qualitative level understand that ripples disappear at large $Ca$, for which the relative importance of surface tension is diminished. A quantitative analysis shows that the critical point gives a diverging wavelength $\lambda \sim 1/(Ca^* - Ca)^{1/3}$, a scenario that is for the first time reported experimentally. 

 
For small $Ca$, our results highlight the importance of the length-scale separation, when one compares the experimental dimple depth with asymptotic solutions. This is of great importance for experiments where the structure of the dimples is used to identify rheological fingerprints \citep{Jacobs1998, Baumchen2010}. From a more general perspective, the used experimental method, digital holographic microscopy, proves to be a powerful tool for the visualisation of the surface profiles of thin films in real-time with height resolution that goes down to the nanometer scale. The methodology can be used in further investigations of films on complex topographies or of fluids with more complex rheology \citep{jalaal2016long}. 

\section*{Acknowledgement}
The authors thank M. Hack, S.G. Huisman, G. Lajoinie, D. Lohse, and N. Balmforth for the
useful discussions. 

\bibliographystyle{jfm}
\bibliography{literature}

\end{document}